# At-Speed Logic BIST for IP Cores

[1] B. Cheon, [1] E. Lee, [2] L.-T. Wang, [3] X. Wen, [4] P. Hsu, [5] J. Cho, [5] J. Park, [4] H. Chao, and [2] S. Wu

[1] Samsung Electronics, Co.  [2] SynTest Technologies, Inc.  [3] Kyushu Institute of Technology
[4] SynTest Technologies, Inc., Taiwan  [5] SynTest Korea, Ltd.

**Abstract**
*This paper describes a flexible logic BIST scheme that features high fault coverage achieved by fault-simulation guided test point insertion, real at-speed test capability for multi-clock designs without clock frequency manipulation, and easy physical implementation due to the use of a low-speed SE signal. Application results of this scheme to two widely used IP cores are also reported.*

## 1. Introduction

Built-In Self-Test for logic circuits or logic BIST implements most of ATE functions on chip, and is especially beneficial for IP cores. Making an IP core a BISTed IP core has the following advantages:

**Simple Test Interface:** Pure logic BIST only needs a start signal to start self-test, a finish signal to indicate its end, and a status signal to show its result. Other test control data is usually small enough to be provided through Boundary-Scan. This completely solves the test access problem when testing an SoC with many embedded cores.
**Better Test Quality:** Logic BIST can easily apply a large number of test patterns. As a result, more defects, either modeled or un-modeled, can be detected. N-detection is also naturally done. In addition, logic BIST makes it easy to conduct at-speed testing for timing-related defects.
**Lower Test Cost:** Logic BIST has most of ATE functions in a chip, so test costs are reduced by less test time, less tester memory requirement, or cheaper tester investment.
**Higher Reliability:** A BISTed IP core can be tested easily after being integrated into a system. Periodic core testing, even with test patterns of relatively low fault coverage, can greatly improve the reliability of the whole system.

Most of logic BIST schemes are based on the STUMPS structure, which applies random patterns generated by a PWG to a full-scan circuit in parallel and compresses the responses into a signature with a MISR. Despite the conceptual simplicity, most of previous logic BIST schemes suffer from the following problems: (1) *Test Frequency Manipulation:* Some logic BIST schemes require all test frequencies satisfy a strict test-oriented relation, which is totally different from functional mode. Not only such a test clocking system is physically difficult to implement but also it fails to realize real at-speed testing; (2) *Performance Concern:* Control points inserted for improving fault coverage add delay to functional paths, thus adversely affecting core performance; (3) *Implementation Difficulty*: Some logic BIST schemes require some test signals, especially the scan enable (SE) signal, to be implemented as high-speed signals that require clock tree synthesis for strict skew management. All these problems limit the use of previous logic BIST scheme in core testing.

## 2. Proposed Logic BIST Scheme

**2.1** General Structure

This paper addresses the above problems with a flexible logic BIST scheme as shown in Fig. 1. It has a TPG block for test pattern generation, an input selector for providing random patterns or top-up ATPG patterns for the core-under-test, a BIST-ready core, an ODC block for output data compression, a clock gating block for generating test clocks from original clocks, and a controller for managing the whole BIST operation [13]. Self-test is started by *Start,* its end is indicated by *Finish,* and its result is shown by *Result.* There is also a standard Boundary-Scan interface, which can be used for loading initial test data or for downloading internal states for fault diagnosis.

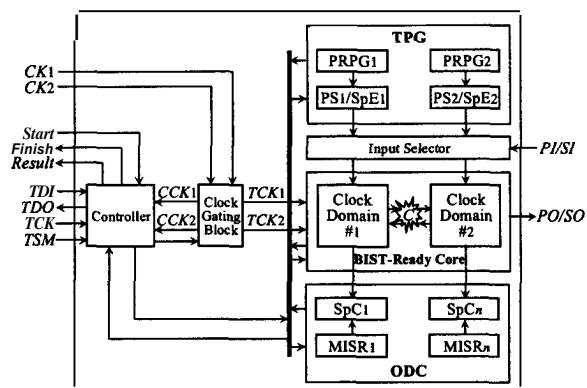

Fig. 1 · · General LBIST structure.

The BIST-ready core is a full-scan circuit with unknown value (*X*) sources properly blocked. In order to directly improve final fault coverage, some observation points are inserted based on the results of fault simulation, instead of observability calculation commonly used in previous logic BIST schemes. In addition, no control point is used in order to meet strict performance requirements for IP cores.

Note that the BIST-ready core has two inter-related clock domains, whose clock skews are usually not aggressively managed. In order to avoid additional design efforts for clock skew management in logic BIST, we use two PRPG-MISR pairs, one for each clock domain, even though they may have the same frequency. In addition, PS1 and PS2 are phase shifters for breaking inter-dependency in raw data streams from PRPGs. Furthermore, space expanders, SpE1 and SpE2, and space compactors, SpC1 and SpC2, are used to reduce the lengths of PRPGs and MISRs.

## 2.2 At-Speed Test Timing Control

This paper proposes a new at-speed test timing control method based on the double-capture scheme. An example is shown in Fig. 2, where test clocks $TCK_1$ and $TCK_2$ are generated by the clock gating block shown in Fig. 1.

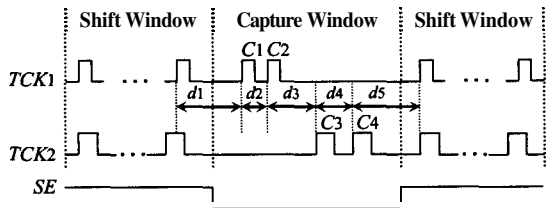

Fig. 2•• At-speed test timing control.

In a capture window, two capture pulses are generated for each clock [1]. The last shift pulse and the first capture pulse are used to create transitions at the outputs of some scan FFs, and the responses to the transitions are caught by the second capture pulse, where $d_2$ and $d_4$ are set based on functional clock frequencies. Thus, real at-speed testing is guaranteed since no test clock frequency manipulation is conducted. In addition, $d_1$ and $d_5$ can be as long as desired, making it possible to use a single and slow scan enable signal (SE) for all clock domains. This significantly eases physical design for logic BIST.

## 2.3 Physical Implementation

In addition to a single and slow SE signal, other techniques are also used to ease physical implementation. This is illustrated in Fig. 3, which only shows two scan chains in Clock Domain #1 and Clock Domain #2.

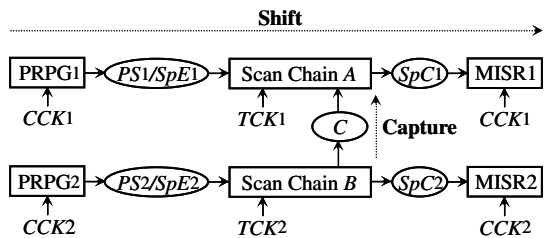

Fig. 3•• Clock skew issues.

In a shift window, a PRPG, a scan chain, and a MISR should operate properly as one shift register. Since the PRPG and the MISR are in a clock domain that is different from the one for the scan chain, timing violations may occur between the PRPG and the scan chain as well as between the scan chain and the MISR. To facilitate physical design, we propose a technique that always makes the clock driving the PRPG and the MISR to be ahead of the clock driving the scan chain in phase. This way, only hold-time violations may occur from the PRPG to the scan chain, while only setup-time violations may occur from the scan chain to the MISR. In this case, the hold-time violations can be corrected with re-timing FFs, while the setup-time violations can be avoided by reducing logic levels from the scan chain to the MISR. This will significantly ease physical design.

In a capture window, as shown in Fig. 2, $d_3$ can be easily adjusted to be larger than the maximal clock skew between the two clock domains. This way, the capture operation can be correctly conducted without adding any state-holding FFs that increase delay on functional paths.

## 3. Experimental Results

The flexible logic BIST scheme has been applied to two commercial CPU IP cores. The results are shown in Table 1.

Table 1•• Experimental Results

|                      | core X        | core Y        |
|----------------------|---------------|---------------|
| Gate Count           | 218.1K        | 633.4K        |
| # of FFs             | 10.3K         | 33.2K         |
| # of Scan Chains     | 100           | 106           |
| Max. Chain Length    | 104           | 345           |
| # of Clock Domains   | 2             | 8             |
| Frequency            | 250MHz        | 330MHz        |
| # of PRPGs           | 2             | 8             |
| PRPG Length          | 19            | 19            |
| # of MISRs           | 2             | 8             |
| MISR Length          | 1: 19 / 1: 99 | 7: 19 / 1: 80 |
| # of Test Points     | 1K (Obv-Only) | 1K (Obv-Only) |
| # of Random Patterns | 20K           | 20K           |
| Fault Coverage 1     | 93.82%        | 93.22%        |
| CPU Time             | 25m43s        | 2h26m48s      |
| Overhead             | 4.4%          | 3.2%          |
| # of Top-Up Patterns | 135           | 528           |
| Fault Coverage 2     | 97.12%        | 97.58%        |

The following techniques were also used in the applications: (1) One PRPG-MISR pair was used for each clock domain since there may be cross-clock-domain logic between any two clock domains. (2) Scan cells were inserted for all PIs and POs to increase delay fault coverage. (3) No space compactor was used between scan outputs and a MISR in order to avoid setup-time violations. This is why there were long MISRs, 99-bit for Core X and 80-bit for Core Y. Such a MISR is generally related to the main and large clock domain which has a larger number of scan chains.

## 4. Conclusions

This paper has described a flexible and pure logic BIST scheme, in which real at-speed testing is realized with easy physical implementation. Successful application of this scheme to two popular IP cores has also been reported.

## Reference

[1] L.-T. Wang, P. Hsu, S. Kao, M. Lin, H. Wang, H. Chao, X. Wen, "A Multiple-Capture DFT System for Detecting or Locating Crossing Clock-Domain Faults During Self-Test or Scan-Test," US *Patent Application,* 20020120896, August 29, 2002.